\documentclass[manuscript]{acmart}

\usepackage{csquotes}
\usepackage{siunitx}
\usepackage{hyperref}

\newcommand{\Xt}[0]{{\textsc{X\textsubscript{t}}}}
\newcommand{\Yt}[0]{{\textsc{Y\textsubscript{t}}}}

\AtBeginDocument{%
  \providecommand\BibTeX{{%
    \normalfont B\kern-0.5em{\scshape i\kern-0.25em b}\kern-0.8em\TeX}}}

\newcommand{\statistics}[4]{($F_{#1, #2}$ = #3, p #4)}
\newcommand{\wald}[3]{($\chi^2(#1) = #2$, p #3)}
\newcommand{\etasquared}[1]{$\eta_{g}^{2}$=#1}
\newcommand{\emmci}[4]{(EMM = #1#4 [#2#4, #3#4])}
\newcommand{\emmcinb}[4]{#1#4 [#2#4, #3#4]}

\usepackage{mathtools}
\usepackage{multirow}
\usepackage{booktabs}
\usepackage{graphicx}

\begin{document}

\title{Understanding Stationary and Moving Direct Skin Vibrotactile Stimulation on the Palm}

\author{Hesham Elsayed}
\affiliation{%
  \institution{TU Darmstadt}
  \city{Darmstadt}
  \country{Germany}
}

\author{Martin Weigel}
\orcid{0000-0002-6171-7369}
\affiliation{%
  \institution{Technische Hochschule Mittelhessen}
  \city{Gießen}
  \country{Germany}
}

\author{Florian Müller}
\affiliation{%
  \institution{LMU Munich}
  \city{Munich}
  \country{Germany}
}

\author{George Ibrahim}
\affiliation{%
  \institution{German University in Cairo}
  \city{Cairo}
  \country{Egypt}
}

\author{Jan Gugenheimer}
\affiliation{%
  \institution{TU Darmstadt}
  \city{Darmstadt}
  \country{Germany}
}

\author{Martin Schmitz}
\affiliation{%
  \institution{Saarland University}
  \city{Saarbrücken}
  \country{Germany}
}

\author{Sebastian Günther}
\affiliation{%
  \institution{TU Darmstadt}
  \city{Darmstadt}
  \country{Germany}
}

\author{Max Mühlhäuser}
\affiliation{%
  \institution{TU Darmstadt}
  \city{Darmstadt}
  \country{Germany}
}

\renewcommand{\shortauthors}{Elsayed et al.}

\begin{abstract}
    Palm-based tactile displays have the potential to evolve from single motor interfaces (e.g., smartphones) to high-resolution tactile displays (e.g., back-of-device haptic interfaces) enabling richer multi-modal experiences with more information. 
    However, we lack a systematic understanding of vibrotactile perception on the palm and the influence of various factors on the core design decisions of tactile displays (number of actuators, resolution, and intensity). In a first experiment (N=16), we investigated the effect of these factors on the users' ability to localize \textit{stationary} sensations. In a second experiment (N=20), we explored the influence of resolution on recognition rate for \textit{moving} tactile sensations.
    Findings show that for \textit{stationary} sensations a 9~actuator display offers a good trade-off and a $3\times3$ resolution can be accurately localized.
    For \textit{moving} sensations, a $2\times4$ resolution led to the highest recognition accuracy, while $5\times10$ enables higher resolution output with a reasonable accuracy. 
\end{abstract}

\ccsdesc[500]{Human-centered computing~Empirical studies in HCI}
\ccsdesc[500]{Human-centered computing~Haptic devices}

\keywords{vibrotactile display, palm-based display, tactile perception, haptics, design guidelines}


\begin{teaserfigure}
 \includegraphics[width=\textwidth]{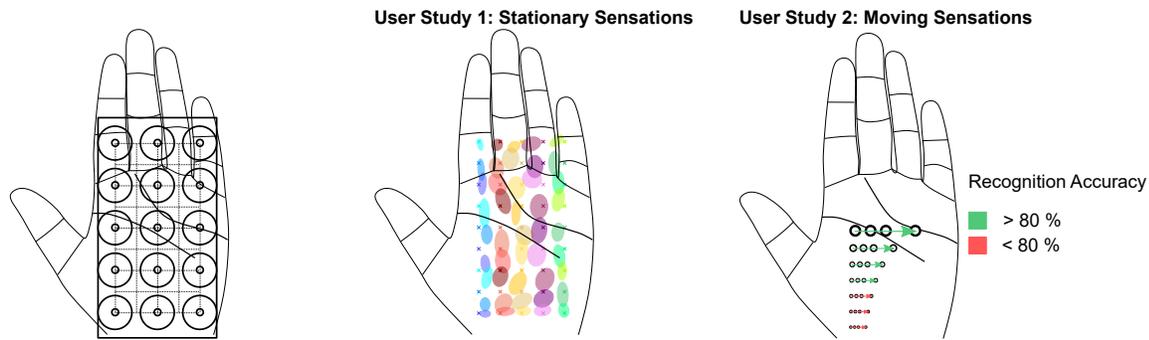}
  \caption{We investigate stationary and moving tactile sensations on the palm that inform the design of haptic interfaces. }
  \label{fig:teaser}
  \Description{The figure shows three hands. The hand on the left has a grid with 15 vibrotactile actuators in a 3x5 grid on it. The hand in the middle shows the error ellipses from study 1 that resulted from localizing stationary sensations on the palm. The hand on the right shows seven moving sensations from study 2 that move towards the right. The sensations vary in length. The first 4 which are the longest resulted in recognition rates over 80\%. The last three, also the shortest, resulted in recognition accuracies under 80\%.}
\end{teaserfigure}

\maketitle

\section{Introduction}
From navigation~\cite{10.1145/2207676.2207734} and skin reading~\cite{Luzhnica:2019:OEV:3290605.3300465} to movement guidance~\cite{Sepp, Marquardt} and haptic learning~\cite{10.1145/3267242.3267271,10.1145/2984511.2984558}, research proposed palm-based vibrotactile displays as a promising output modality for a diverse set of use-cases.
These interfaces are especially applicable in situations where multi-modal interaction is beneficial, where interaction with video and audio displays is infeasible or not recommended -- e.g., while driving or riding a bike -- or where subtle interaction is required -- e.g., while holding a conversation. Research has started to explore how to encode information via vibrotactile patterns, that are either \textit{stationary} at a fixed location or \textit{moving} over time between various spatial locations.

One of the most promising locations for applying vibrotactile patterns is the palm due to high sensitivity and the frequency with which we use our hands to interact with the environment. Fundamental to the design of palm-based tactile displays is their resolution. Research has investigated the spatial acuity (2-point discrimination and point localization)~\cite{SAPT} of the human body to touch at various locations, including the palm. However, prior work has shown that vibrotactile sensations have a different spatial acuity than touch~\cite{10.1145/3432189}. This difference occurs because vibrations (1) activate the Pacinian corpuscle mechanoreceptors~\cite{ROWIN2007343} with larger receptive fields and (2) propagate larger distances on the skin~\cite{Cholewiak2003}, making them harder to localize. Thus, given that vibration is a commonly used modality in haptics research, we currently lack information on the perception of \textit{stationary} and \textit{moving} tactile sensations. This information is necessary for the design and usage of palm-based tactile displays.

In this paper, we contribute important insights on \textit{stationary} and \textit{moving} vibrotactile sensations on the palm. In a first experiment, we investigated the influence of the layout and number of actuators, intensity, and resolution on the localization error and perception of real and phantom \textit{stationary sensations}. Based on the results, we determined that a $3\times3$ resolution can be accurately localized on the palm. Nine vibration motors with a bigger (approximately 2:1 ratio) spacing along the length of the palm than the width should be used. We further observed that increasing the number of actuators to 15 resulted in a significant increase in correct perception of phantom sensations at a single location.

In a second experiment, we investigated \textit{moving} sensations. In particular, we explored the influence of the resolution of the display and direction of movement on the recognition accuracy and reaction time of users. We observed that a $2\times4$ resolution can be used for \textit{accurate} interactions with \textit{moving} sensations (recognition accuracy > 95\%). While a $5\times10$ resolution can be used where more \textit{expressive} tactile sensations are required (recognition accuracy > 85\%). Based on the findings of our two experiments, we contribute a set of design guidelines for vibrotactile output on the palm.

Taken together, the main contributions of this paper are:
\begin{enumerate}
\item Findings from a controlled user study investigating perception of \textit{stationary} tactile sensations.
\item Findings from a controlled user study investigating perception of \textit{moving} tactile sensations.
\item A set of design guidelines based on our findings to improve future vibrotactile displays on the palm.
\end{enumerate}
\section{Related Work}
This work relates to prior research in measuring the spatial acuity of the palm, leveraging vibrotactile illusions, and work on \textit{stationary} and \textit{moving} tactile sensations on the palm.

\subsection{Spatial Acuity of the Palm}
Various aspects of vibrotactile perception have been investigated on the body, for example temporal aspects of perceiving vibrations as distinct~\cite{doi:10.2466/pms.1989.68.1.288}, the effect of the number of actuators on perceived intensity~\cite{sfpi}, and parameters effecting perception of patterns on the body~\cite{PMID:232572}. In the following, work related to vibrotactile spatial acuity of the palm is discussed.
Research~\cite{Weinstein,weber1978sense} has shown that the spatial acuity of the sense of touch varies across the body, with the  palm being among the most sensitive body parts, superseded only by the fingertips~\cite{SAPT}. Ever since the seminal work of Weber~\cite{weber1834subtilitate} on touch, spatial acuity has been measured by two-point discrimination thresholds. These refer to the minimum distance required between two simultaneous stimuli for them to be perceived as distinct. However, this wealth of knowledge cannot be used to inform the core design decisions of vibrotactile displays (e.g., spacing of actuators), as prior work has shown that spatial acuity of the body to vibrotactile stimulation is fundamentally different than touch~\cite{10.1145/3432189}. This is mainly due to vibrations activating mechanoreceptors with a larger receptive field (Pacinian corpuscle)~\cite{ROWIN2007343} and the fact that vibrations propagate for larger distances on the skin~\cite{Cholewiak2003}.

To overcome this, recent research has investigated vibrotactile perception on major body locations, e.g., forearm, upper arm, thigh, stomach,  back, and leg~\cite{10.1145/3432189}. Findings show that the spatial acuity of the body to vibrotactile stimulation follows a similar trend to touch regarding the sensitivity of body locations, however, with considerably different absolute values. To the best of our knowledge, a systematic investigation of vibrotactile perception on the palm remains unexplored. The palm is a prime location for vibrotactile feedback due to high sensitivity and the frequency with which we use our hands to interact with the environment. Therefore, this work is concerned with measuring vibrotactile perception on the palm---localization error of \textit{stationary} vibrations and recognition rate of \textit{moving} sensations across resolutions---to inform the main decisions associated with the design and usage of palm-based tactile displays. Although several factors affect vibrotactile perception, e.g. body site, choice of actuator, and actuator mounting conditions, prior work has identified that the main factors that affect localization at a particular body site to be the number and spacing of actuators~\cite{sofia}.

\subsection{Vibrotactile Illusions}
Tactile illusions have proven to be useful in HCI applications due to their ability to generate sensations where no physical actuator is present, thus rendering high resolution spatial vibrotactile stimuli using a low resolution grid of actuators. The three most common tactile illusions are \emph{phantom sensations}~\cite{DSAlles,Park:2018:TIT:3173574.3173832}, \emph{cutaneous rabbit}~\cite{Geldard1977SensoryS,10.1145/1647314.1647381,Spelmezan:2009:TMI:1518701.1519044,McDaniel}, and \emph{apparent tactile motion}~\cite{Sherrick1966ApparentHM,Kirman1974TactileAM,burt}. Although all of these illusions can generate robust sensations, only \emph{phantom sensations} have the ability to produce \textit{stationary} and \textit{moving} sensations. Therefore, in this work we use phantom sensations to increase the resolution of palm-based vibrotactile displays.

Also known as the funneling illusion, phantom sensations refer to the illusion where the perceived location of a vibration is controlled by varying intensity between two (1D phantom sensations) or more (2D phantom sensations) neighbouring vibrotactile actuators~\cite{DSAlles,Park:2018:TIT:3173574.3173832}. Tactile interfaces have leveraged phantom sensations to generate expressive patterns. Mango~\cite{Schneider:2015:TAD:2807442.2807470}, an authoring tool for creating vibrotactile patterns, uses direct manipulation of phantom sensations for designing and rendering 2D patterns. In Tactile Brush~\cite{Israr:2011:TBD:1978942.1979235}, an algorithm is proposed and validated that uses phantom sensations and apparent tactile motion to generate high resolution 2D vibrotactile strokes.

\subsection{Stationary Sensations on the Palm}
There is a vast literature on work that has leveraged vibrotactile sensations on the body, e.g., on the  hand~\cite{10.1145/2984511.2984558,Sepp,Lehtinen:2012:DTG:2380116.2380173,Park:2018:TIT:3173574.3173832,Gunther:2018:EAV:3197768.3201568}, wrist~\cite{Lee:2010:BAP:1753326.1753392,Lee:2015:IIT:2702123.2702530, Cauchard:2016:ADE:2858036.2858046,Liao:2016:EEA:2984511.2984522}, forearm~\cite{Zhao:2018:CTS:3173574.3173966,Luzhnica:2017:PVD:3123021.3123029,Luzhnica:2019:OEV:3290605.3300465,Reinschluessel:2018:VSN:3170427.3188549,Pfeiffer:2014:LMG:2582051.2582099,Schonauer:2012:MMG:2388676.2388706}, upperarm~\cite{Stratmann:2018:EVP:3205873.3205874,KBark,Alvina:2015:OTC:2702123.2702341}, back~\cite{Israr:2011:TBD:1978942.1979235,Back,AHB}, stomach~\cite{Kruger:2018}, thigh~\cite{Spelmezan:2009:TMI:1518701.1519044} and leg~\cite{Chen:2018:ETP:3206505.3206511}. This section outlines the most relevant related work that focuses on \textit{stationary} sensations on the palm. 

\subsubsection{Layout \& Number of Actuators}
Palm-based tactile displays have taken many shapes, e.g.: spherical handles~\cite{10.5555/1703775.1704019}, square/diamond arrangements~\cite{5945469}, and grids~\cite{Alvina:2015:OTC:2702123.2702341, Park:2018:TIT:3173574.3173832,10.1145/2207676.2207734,10.1145/1622176.1622198}. The number of actuators ranged from one~\cite{10.1145/2702123.2702396} to 30~\cite{Borst} with inter-actuator spacing varying depending on the display. In our work, we systematically investigate the influence of the layout and number of actuators on the perception of \textit{stationary} sensations.

\subsubsection{Resolution}
Typically, prior work used a resolution that is defined by the number of real actuators on the palm~\cite{Alvina:2015:OTC:2702123.2702341,10.1145/1622176.1622198,5945469}. However, approaches also exist that extend the resolution by leveraging phantom sensations~\cite{Park:2018:TIT:3173574.3173832,10.5555/1703775.1704019}. It is unclear, what the maximum resolution is, where accurate localization of \textit{stationary} sensations is still possible. In our work, we systematically investigate the interdependency between localization accuracy and resolution and the extent to which phantom sensations can overcome the limits of physical resolution.

\subsection{Moving Sensations on the Palm}
In addition to \textit{stationary} sensations, many approaches in the literature were introduced that focus on evaluating the utility of \textit{moving} tactile sensations~\cite{7177686,5651759,6088602,6181408,6343763,Alvina:2015:OTC:2702123.2702341,10.1007/978-3-319-42321-0_5}.

Prior work has used many different resolutions for generating \textit{moving} sensations on the palm, ranging between grids of four~\cite{7177686} to 12 actuators~\cite{5651759,6088602,6181408,6343763}. Recognition rates of \textit{moving} sensations vary from 70-80\%~\cite{Alvina:2015:OTC:2702123.2702341,6181408,6343763} up to above 90\%~\cite{Alvina:2015:OTC:2702123.2702341,5651759,6088602}, depending on the number of patterns, actuators, and inter-actuator spacing. While most work focuses on discriminating between a set of distinct patterns~\cite{7177686,5651759,6088602,6181408,6343763,Alvina:2015:OTC:2702123.2702341}, some approaches were aimed at providing a display for continuous sensations~\cite{10.1007/978-3-319-42321-0_5}. 

In this paper, we aim to systematically investigate the influence of resolution on the vibrotactile perception (recognition rate) of users. This information is missing in the literature and is critical for the usage and design of palm-based vibrotactile displays.

\section{User Study 1: Stationary Tactile Sensations}
\begin{figure*}
  \centering
  \includegraphics[width=0.75\textwidth]{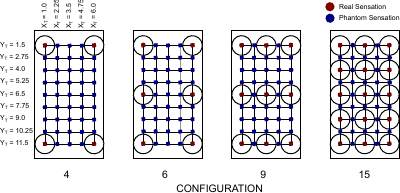}
  \caption{Independent variables of user study 1: \textsc{configuration}, target X (\textsc{X\textsubscript{T}}), and target Y (\textsc{Y\textsubscript{T}}).}
  \label{fig:study1}
\end{figure*}

Spatial acuity of touch is different from that of vibrations~\cite{10.1145/3432189}. To get a better understanding of how we perceive vibrotactile \textit{stationary} sensations on the palm, this user study aimed to answer the following research questions:

\begin{description}
    \item[RQ1] How does the choice of configuration (number and layout of actuators) influence the localization accuracy and perception of vibrations? Prior work used many different configurations. It is unclear, how the choice of configuration effects the localization accuracy and perception of vibrations on the palm.
    
    \item[RQ2] How does the intensity of vibration affect the localization accuracy and perception of vibrations? We hypothesized that higher intensity vibrations increase localization error due to a wider area of propagation on the skin.
    
    \item[RQ3] The number of actuators and phantom sensations can increase the tactile resolution. How does the choice of resolution influence the localization accuracy? Prior work has utilized one resolution per configuration. It is unclear, how the choice of resolution influences the localization accuracy.
\end{description}

In this section, we describe our study design, the procedure, our participants, apparatus, dependent variables, and data analysis methods.

\subsection{User Study Design}
Throughout the user study we varied the following four independent variables:
\begin{description}
    \item[\textsc{Configuration}:] The number of vibration motors in the palm-based tactile display. \textsc{Configuration} has 4 levels: 4, 6, 9, and 15 vibration motors. Figure~\ref{fig:study1} illustrates the placement of the vibration motors in the grid. We chose grids as they are most frequently used~\cite{Alvina:2015:OTC:2702123.2702341, Park:2018:TIT:3173574.3173832,10.1145/2207676.2207734,10.1145/1622176.1622198} and because they enable the use of 2D phantom sensations~\cite{Park:2018:TIT:3173574.3173832}. The number of actuators was systematically varied by adding rows and columns.
    
    \item[\textsc{Intensity}:] The intensity of the vibrations with 2~levels: 0.5 ($0.5 \cdot Amplitude_{max}$ of the EAI C2 tactor) and 1.0 (vibrations with maximum amplitude of the EAI C2 tactor). All vibrations were performed at a fixed frequency (\SI{200}{\Hz}~\cite{Park:2018:TIT:3173574.3173832}) and lasted one second.
    
    \item[\textsc{\Xt}:] The stimulus position of the tactile sensation on the x-axis. \textsc{\Xt} has 5~levels translating to 5~columns as shown in \autoref{fig:study1}. We chose 5~columns along the width of the palm based on prior work~\cite{Park:2018:TIT:3173574.3173832}.
    
    \item[\textsc{\Yt}:] The stimulus position of the tactile sensation on the y-axis. \textsc{\Yt} has 9~levels translating to 9~rows as shown in \autoref{fig:study1}. We chose 9~rows based on related work~\cite{Park:2018:TIT:3173574.3173832} and to keep the same spacing as the x-axis. Depending on the \textsc{configuartion}, \textsc{\Xt}, and \textsc{\Yt}, the stimulus was either a real or phantom sensation. For generating phantom sensations, we used the same approach as Park and Choi~\cite{Park:2018:TIT:3173574.3173832} as described in the next section.
    
\end{description}
The user study contained a total of 360 ($4 \times 2 \times 5 \times 9$) conditions and followed a within subjects study design. We used an $8 \times 8$ balanced latin square to counterbalance the variables \textsc{configuration} and \textsc{intensity}. For each combination of these independent variables, participants experienced 45~vibration locations ($5\times9$). The order of these locations was randomized and each location was repeated only once, resulting in a total of 360~trials per participant.

\subsection{Generation of Phantom Sensations}

To generate 1D and 2D phantom sensations we used an algorithm~\cite{Park:2018:TIT:3173574.3173832} that controls the intensities of four actuators arranged in a grid to interpolate between them. The vibration intensity of an actuator i is calculated using the following equation.

\[
Intensity_i = Intensity_{target}( 1 - \frac{d_i^x}{D^x})(1 - \frac{d_i^y}{D^y})
\]

where $d_i^x$ and $d_i^y$ are the horizontal and vertical distances from the target point to actuator i, respectively. $Intensity_{target}$ is the target intensity of the phantom sensation. The algorithm is used to calculate the intensities of all four actuators involved in the generation of a 2D phantom sensation. In case of 1D phantom sensations, the actuators not involved in the generation of the sensation are inactive as $d^x = D^x$ or $d^y = D^y$.

\begin{figure*}
  \centering
  \includegraphics[width=\textwidth]{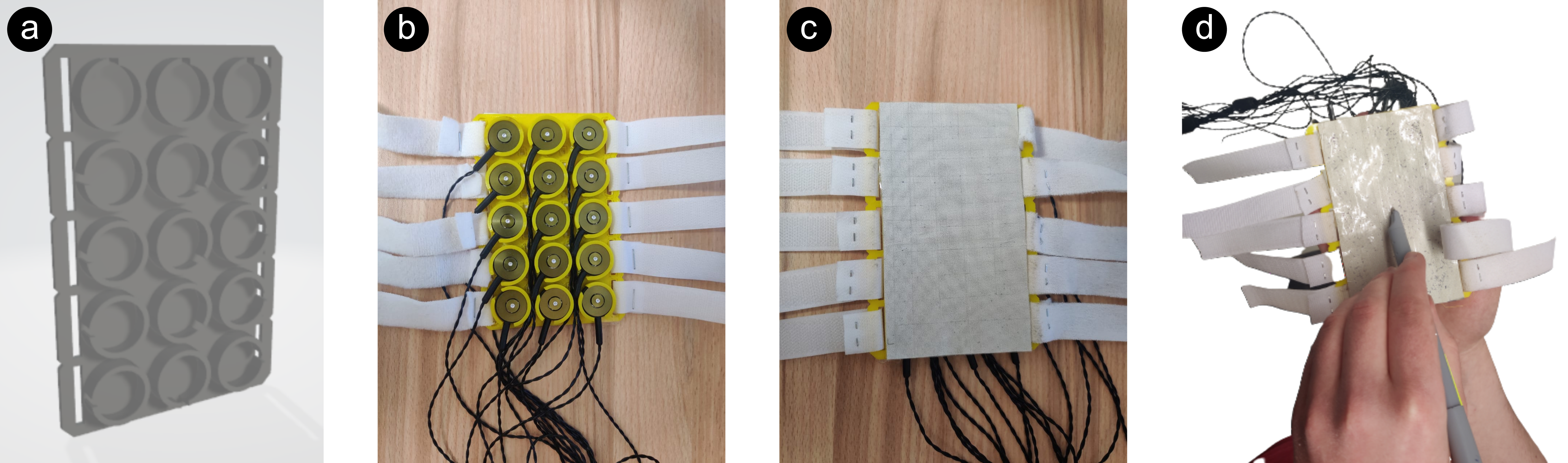}
  \caption{Apparatus used in user study 1: (a) 3D model of vibrotactile grid, (b\&c) the grid from both sides, and (d) the smartpen used for input on the digital paper. }
  \label{fig:apparatus}
\end{figure*}

\subsection{Apparatus}
The prototype consisted of three parts: vibrotactile actuators, 3D printed case with smartpen paper attached to the backside, and a smartpen.
We printed our prototype grids with a Prusa i3 MK3S+ using thermoplastic polyurethane (TPU) filament. The grids contained holders for attaching EAI C2 vibration motors~\cite{eai}. Furthermore, the EAI C2 tactor featured a small contact spot (7.6 mm diameter) enclosed within a larger rigid cylindrical housing (30.5 mm diameter) that prevented the spread of vibration~\cite{sofia}. For input, we used a Neo Smartpen M1. A clear coating on top of the NeoLab paper prevented abrasion and visible marks of prior inputs. Five bands of Velcro tape on the sides of our prototypes allowed attachment on participants' palms while ensuring that all vibration motors are in contact with the skin. \autoref{fig:apparatus} shows a prototype used in our experiment.

The prototype was connected to an i7 dual core 3.6~GHz 16~GB RAM desktop PC, which ran the software used in our user study. The software consisted of a C\# project that received data from the smartpen over Bluetooth and controlled the vibration motors over USB.

\subsection{Procedure}
After obtaining informed consent from the participants, we collected their demographic data. Then, we explained the task and provided a brief overview of the procedure. The task was to indicate the location of the vibration using the digital pen. 

At the beginning, we asked participants to wear noise cancelling headphones playing white noise to prevent the sound of vibrations influencing their answers. Each trial started with the participant in a seated position with their hands resting on the armrest of the chair and their palm side up. All participants were right-handed and hence wore the grid on their left palm and held the pen in their right hand. After experiencing a vibration, the experimenter asked the participant if the vibration was at one location or more than one location. The experimenter explained to the participants that they should indicate after the number of perceived points after each trial before starting the experiment. Communication about the number of perceived points was accomplished using hand gestures so that participants were not required to remove the headphones after each trial. All stimuli were targeted at one location using real and phantom sensations. Participants were instructed to indicate a location in the middle if they perceived more than one vibration. Participants were further instructed to wait until the vibration was over before using the pen.

Participants took a break (approximately five minutes long) every 90~vibrations. This resulted in four breaks and a total duration of about 60~minutes for conducting the user study. 

\subsection{Dependent Variables}
The following dependent variables were measured:

\begin{description}
    \item[Euclidean Distance:] The euclidean distance between the perceived and target location of the vibration. 
    
    \item[X Deviation:] The deviation on the x-axis between the target and perceived location.
    
    \item[Y Deviation:] The deviation on the y-axis between target and perceived location.
    
    \item[Accuracy:] The accuracy of localizing target location. A response is considered correct when the closest point is the target location.
    
    
    
    \item[Number of perceived points:] The number of vibrations perceived by the user (binary: either one point or two or more points) 
\end{description}

\subsection{Participants}
We recruited 16 right-handed participants (12 male and 4 female) aged between 20 and 32 years old ($\mu = 23.19$, $\sigma = 2.88$). None of the participants had prior experience with vibrotactile feedback beyond the everyday use of smartphones and game controllers and no sensory processing disorders were reported by our participants. 

\subsection{Data Analysis}
After visually confirming that the data follows a normal distribution, we used four-way repeated measures (RM) ANOVAs with the factors \textsc{configuration}, \textsc{intensity}, \textsc{\Xt}, and \textsc{\Yt} to compute the F-score and p-value of main and interaction effects. Where Mauchly's test indicated a violation of the assumption of sphericity, we used the Greenhouse Geisser method. We further report the generalized eta-squared $\eta_{g}^{2}$ as an estimate of the effect size and use Cohen’s suggestions to classify the effect size as small, medium or large~\cite{Cohen1988}. If significant effects were found, we used pairwise t-tests with Tukey adjustment for post-hoc analysis. Furthermore, we report the estimated marginal mean (EMM) with 95\% confidence intervals as proposed by Searle et al.~\cite{searle1980population}.
\begin{figure*}
  \centering
  \includegraphics[width=\textwidth]{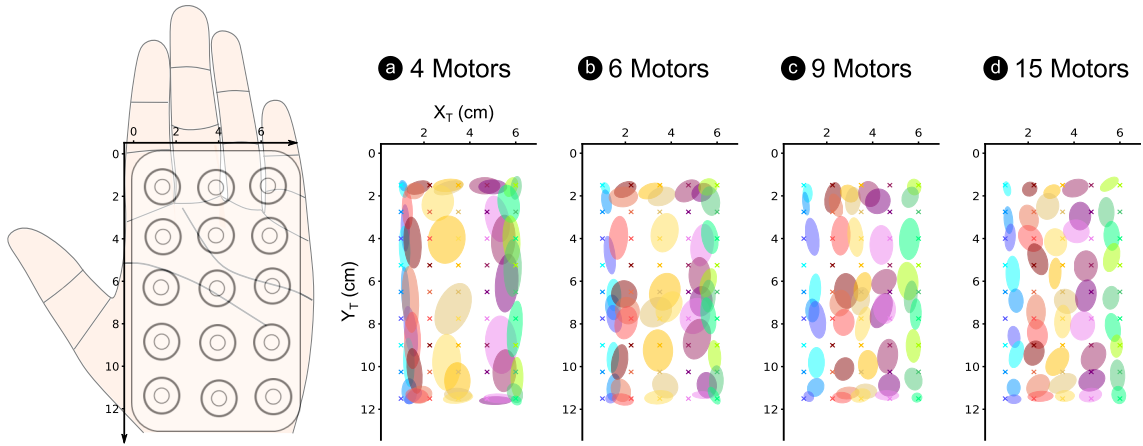}
  \caption{Error ellipses for (a) \textsc{4 motor}, (b) \textsc{6 motor}, (c) \textsc{9 motor}, and (d) \textsc{15 motor configurations} based on the mean and covariance of point clouds at the different locations defined by \Xt{} and \Yt{} ($\sigma = .5$).}
  \label{fig:ellipses}
\end{figure*}

\section{User Study 1: Results}
In the following, we report the results of our first user study as detailed in the prior section. We label key observations with \textbf{[MF-\#]} for main effects and \textbf{[IF-\#]} for interaction effects. The recorded location data is visualized in \autoref{fig:ellipses}.

\begin{figure*}
  \centering
  \includegraphics[width=\textwidth]{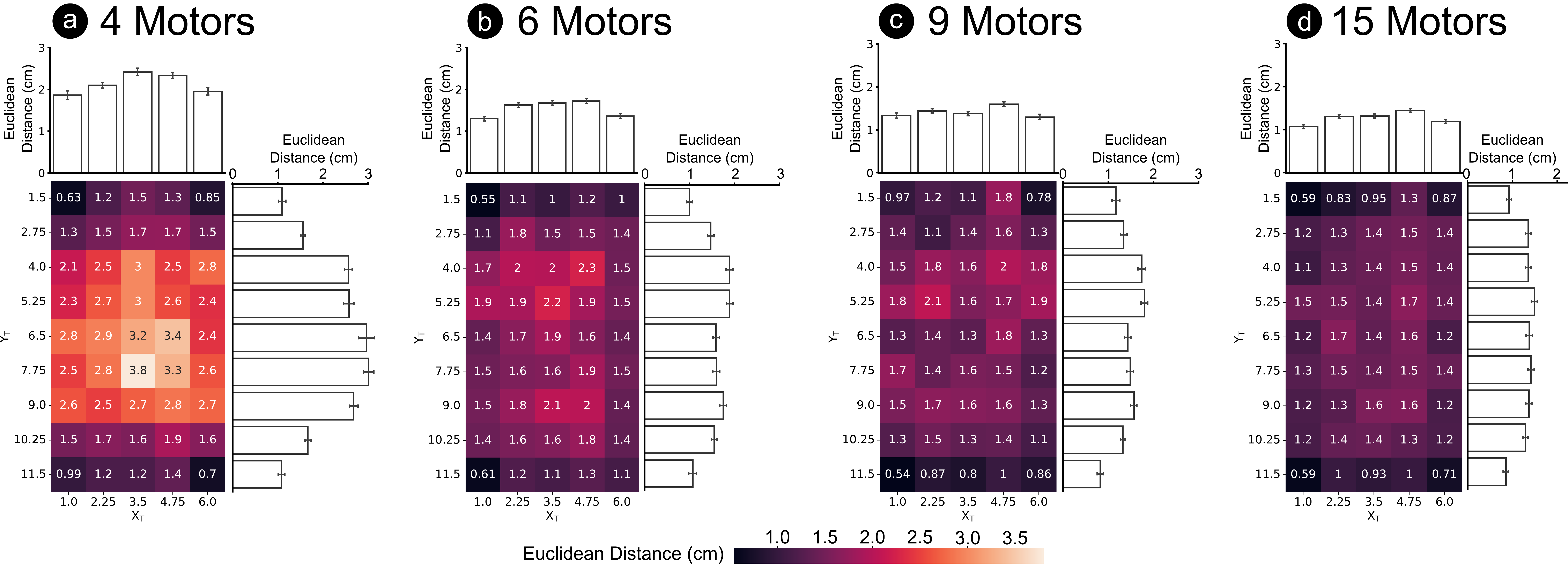}
  \caption{Illustration of the influence of the factors \textsc{configuration} (a = \textsc{4 motor}, b = \textsc{6 motor}, c = \textsc{9 motor}, and d = \textsc{15 motor}), \Xt, \Yt and the interactions \textsc{configuration:\Xt} and \textsc{configuration:\Yt} on the euclidean distance. Error bars are the standard error. }
  \label{fig:l2norm}
\end{figure*}

\subsection{Euclidean Distance}
To get an understanding of the influence of the factors on the general localization accuracy of users, we measured the euclidean distance between target locations (\Xt{} and \Yt) and perceived locations. \autoref{fig:l2norm} shows the results. Our analysis revealed no significant four-way and three-way interaction effects. We found two significant two-way interactions. In the following, we report the main effects and significant interaction effects.

\subsubsection{\textsc{configuration}}
\textbf{[MF-1]} The analysis showed a significant \statistics{1.54}{23.09}{45.53}{< .001} main effect of the factor \textsc{configuration} on the euclidean distance with a medium (\etasquared{.10}) effect size. We found that the \textsc{15 motors configuration} \emmci{1.28}{1.16}{1.40}{cm} resulted in the lowest errors, followed by the \textsc{9~motors configuration} \emmci{1.41}{1.30}{1.52}{cm}, the \textsc{6~motors configuration} \emmci{1.54}{1.45}{1.63}{cm}, and finally the \textsc{4~motors configuration} \emmci{2.12}{1.93}{2.31}{cm}. Post-hoc tests confirmed significant differences between the \textsc{4~motors configuration} and all other \textsc{configurations} (p < .001) and between the \textsc{6~motors} and \textsc{15~motors configurations} (p < .001).

\subsubsection{\textsc{intensity}}
We could not find a significant main effect for the factor \textsc{intensity} \statistics{1}{15}{1.16}{> 0.05} on the euclidean distance between 0.5 \textsc{intensity} \emmci{1.60}{1.50}{1.71}{cm} and 1.0 \textsc{intensity} \emmci{1.57}{1.49}{1.65}{cm}.

\subsubsection{\Xt}
\textbf{[MF-2]} Our analysis revealed a significant \statistics{2.95}{44.23}{11.13}{< .001} main effect of the factor \Xt{} on the euclidean distance with a small (\etasquared{.02}) effect size. The euclidean distance was lowest for \Xt{} at the edges of the vibration grid, \Xt = 1.0 \emmci{1.40}{1.25}{1.54}{cm}, \Xt = 6.0 \emmci{1.45}{1.29}{1.62}{cm} and higher for locations in the middle: \Xt = 2.25 \emmci{1.62}{1.53}{1.70}{cm}, \Xt = 3.5 \emmci{1.70}{1.57}{1.82}{cm}, and \Xt = 4.75 \emmci{1.77}{1.68}{1.87}{cm}.

\subsubsection{\Yt}
\textbf{[MF-3]} The analysis revealed a significant \statistics{4.42}{66.26}{41.98}{< .001} main effect of the factor {\Yt} on the euclidean distance with a medium (\etasquared{.12}) effect size. Similar to \Xt, the lowest euclidean distances were at the top and bottom edges of the vibration grid: \Yt = 1.5 \emmci{1.06}{0.88}{1.23}{cm} and \Yt = 11.5 \emmci{0.97}{0.83}{1.12}{cm}. Rows in the middle showed higher euclidean distances: \Yt = 2.75 \emmci{1.43}{1.33}{1.54}{cm}, \Yt = 4.0 \emmci{1.88}{1.75}{2.01}{cm}, \Yt = 5.25 \emmci{1.94}{1.80}{2.08}{cm}, \Yt = 6.5 \emmci{1.84}{1.66}{2.01}{cm}, \Yt = 7.75 \emmci{1.88}{1.71}{2.04}{cm}, \Yt = 9.0 \emmci{1.83}{1.70}{1.96}{cm}, and \Yt = 10.25 \emmci{1.46}{1.35}{1.57}{cm}.

\subsubsection{\textsc{configuration : \Xt}}
\textbf{[IF-1]} The analysis showed a significant \statistics{5.74}{86.07}{2.30}{< .05} interaction effect between the factors \textsc{configuration} and \Xt{} with a small effect size (\etasquared{.01}). The euclidean error depended on the combination of \textsc{configuration} and \Xt, with \textsc{configurations} using a lower number of actuators (4 and 6) showing significant differences (p < .05) as \Xt{} changes, and \textsc{configurations} using a higher number of actuators (9 and 15) showing comparable performance (p > .05) across changing \Xt.

\subsubsection{\textsc{configuration : \Yt}}
\textbf{[IF-2]} The analysis revealed a significant \statistics{24}{360}{9.12}{< .001} interaction effect between the factors \textsc{configuration} and \Yt{} with a medium effect size (\etasquared{.06}). The euclidean distance was significantly lower at the top and bottom edges of the grid for all \textsc{configurations}. However, for \textsc{configurations} using a lower number of actuators (4, 6, and 9), significant differences (p < .05) were observed in the range 1.5 < \Yt < 11.5, whereas the \textsc{configuration} with the highest number of actuators (15) showed comparable performance (p > .05).

\subsection{X Deviation}
To get a better understanding of the influence of the factors on users' ability to localize sensations along the width of the palm, we analyzed the deviations in the x-axis between the target and perceived locations.

\subsubsection{\textsc{Configuration}}
\textbf{[MF-4]} The analysis revealed a significant \statistics{2}{30.05}{21.06}{< .001} main effect of the factor \textsc{configuration} on the recorded X deviation with a small (\etasquared{.02}) effect size. We found that the \textsc{9 motors configuration} \emmci{0.64}{0.58}{0.71}{cm} resulted in the lowest errors, followed by the \textsc{15~motors configuration} \emmci{0.66}{0.58}{0.74}{cm}, the \textsc{6~motors configuration} \emmci{0.77}{0.70}{0.83}{cm}, and finally the \textsc{4~motors configuration} \emmci{0.86}{0.78}{0.94}{cm}. Post-hoc tests confirmed significantly decreasing X~deviation going from a lower number of actuators (4,6) to a higher number of actuators (9,15) (p < .001). X~deviation was comparable for the \textsc{configuration} pairs (4,6) and (9,15).

\subsubsection{\textsc{Intensity}}
The analysis showed no significant \statistics{1}{15}{0.26}{> .05} main effect of the factor \textsc{intensity} on the X~deviation. \textsc{0.5 intensity} \emmci{0.73}{0.66}{0.79}{cm} showed comparable X deviation to \textsc{1.0 intensity} \emmci{0.74}{0.67}{0.80}{cm}.

\subsubsection{\Xt}
\textbf{[MF-6]} The analysis revealed a significant \statistics{2.84}{42.61}{23.51}{< .001} main effect of the factor \Xt{} on the X~deviation with a medium (\etasquared{.10}) effect size. The X~deviation was lowest for \Xt{} at the left \Xt = 1.0 \emmci{0.48}{0.36}{0.60}{cm} and right \Xt = 6.0 \emmci{0.50}{0.37}{0.64}{cm} edges of the vibration grid. Locations in the middle showed higher X deviation: \Xt = 2.25 \emmci{0.83}{0.78}{0.88}{cm}, \Xt = 3.5 \emmci{0.84}{0.71}{0.98}{cm}, and \Xt = 4.75 \emmci{1.00}{0.93}{1.07}{cm}. Post-hoc confirmed signficantly rising X deviation going from the edges (\Xt = 1.0, \Xt = 6.0) towards the middle (\Xt = 2.25, \Xt = 3.5, \Xt = 4.75) (p < .001).

\subsubsection{\Yt}
We could not find a significant \statistics{4.67}{70.09}{1.71}{> .05} main effect of the factor \Yt{} on the recorded X~deviation. Comparable values for X~deviation were observed across all levels:
\Yt = 1.5 \emmci{0.68}{0.62}{0.75}{cm},
\Yt = 2.75 \emmci{0.70}{0.64}{0.75}{cm},
\Yt = 4.0 \emmci{0.75}{0.68}{0.82}{cm},
\Yt = 5.25 \emmci{0.75}{0.69}{0.82}{cm},
\Yt = 6.5 \emmci{0.79}{0.69}{0.89}{cm},
\Yt = 7.75 \emmci{0.74}{0.65}{0.83}{cm},
\Yt = 9.0 \emmci{0.73}{0.65}{0.81}{cm},
\Yt = 10.25 \emmci{0.74}{0.65}{0.83}{cm}, and \Yt = 11.5 \emmci{0.70}{0.62}{0.79}{cm}.

\subsection{Y Deviation}
To get a better understanding of the influence of the factors on users' ability to localize sensations along the length of the palm, we analyzed deviations in the y-axis between target and perceived locations.

\subsubsection{\textsc{Configuration}}
\textbf{[MF-7]} The analysis showed a significant \statistics{1.66}{24.87}{37.23}{< .001} main effect of the factor \textsc{configuration} on the Y~deviation with a medium (\etasquared{.09}) effect size. The \textsc{4 motor configuration} resulted in the highest Y deviation \emmci{1.75}{1.56}{1.93}{cm}, followed by the \textsc{6 motor configuration} \emmci{1.15}{1.07}{1.24}{cm}, the \textsc{9 motor configuration} \emmci{1.11}{1.01}{1.21}{cm}, and finally the \textsc{15 motor configuration} \emmci{0.94}{0.84}{1.04}{cm}. Post-hoc tests confirmed significant differences between the \textsc{4 motor configuration} and all other \textsc{configurations} (p < .001), between 6 and 15 (p < .001), and between 9 and 15 (p < .05).

\subsubsection{\textsc{Intensity}}
We could not find a significant main effect for the factor \textsc{intensity} \statistics{1}{15}{3.85}{> 0.05} on the Y deviation between \textsc{0.5 intensity} \emmci{1.27}{1.19}{1.34}{cm} and \textsc{1.0 intensity} \emmci{1.21}{1.15}{1.27}{cm}.

\subsubsection{\Xt}
We could not find a significant main effect for the factor {\Xt} \statistics{3.05}{45.81}{1.22}{> 0.05} on the Y deviation. We observed comparable Y deviation for \Xt = 1.0 \emmci{1.19}{1.09}{1.29}{cm}, \Xt = 2.25 \emmci{1.21}{1.13}{1.29}{cm},
\Xt = 3.5 \emmci{1.27}{1.20}{1.34}{cm},
\Xt = 4.75 \emmci{1.27}{1.20}{1.35}{cm}, and \Xt = 6.0 \emmci{1.24}{1.13}{1.36}{cm}.

\subsubsection{\Yt}
\textbf{[MF-8]} The analysis showed a significant \statistics{4.22}{63.30}{44.96}{< .001} main effect of the factor \Yt{} on the Y~deviation with a large (\etasquared{.16}) effect size. Y~deviation was lowest at the top \Yt = 1.5 \emmci{0.62}{0.43}{0.81}{cm} and the bottom \Yt = 11.5 \emmci{0.48}{0.33}{0.62}{cm} rows of the grid. Rows in the middle showed higher Y~deviation values: \Yt = 2.75 \emmci{1.11}{0.98}{1.23}{cm}, \Yt = 4.0 \emmci{1.58}{1.45}{1.72}{cm}, \Yt = 5.25 \emmci{1.63}{1.49}{1.77}{cm}, \Yt = 6.5 \emmci{1.48}{1.30}{1.66}{cm}, \Yt = 7.75 \emmci{1.57}{1.42}{1.72}{cm}, \Yt = 9.0 \emmci{1.54}{1.42}{1.67}{cm}, and \Yt = 10.25 \emmci{1.12}{1.02}{1.23}{cm}. Post-hoc tests confirmed significantly rising Y deviation from the top and bottom (\Yt = 1.5, \Yt = 11.5), to the next two rows (\Yt = 2.25, \Yt = 10.25) (p < .01), and between \Yt = 2.25, \Yt = 10.25 and all other rows in the middle (p < .05).

\subsection{Accuracy}
\begin{figure*}
  \centering
  \includegraphics[width=\textwidth]{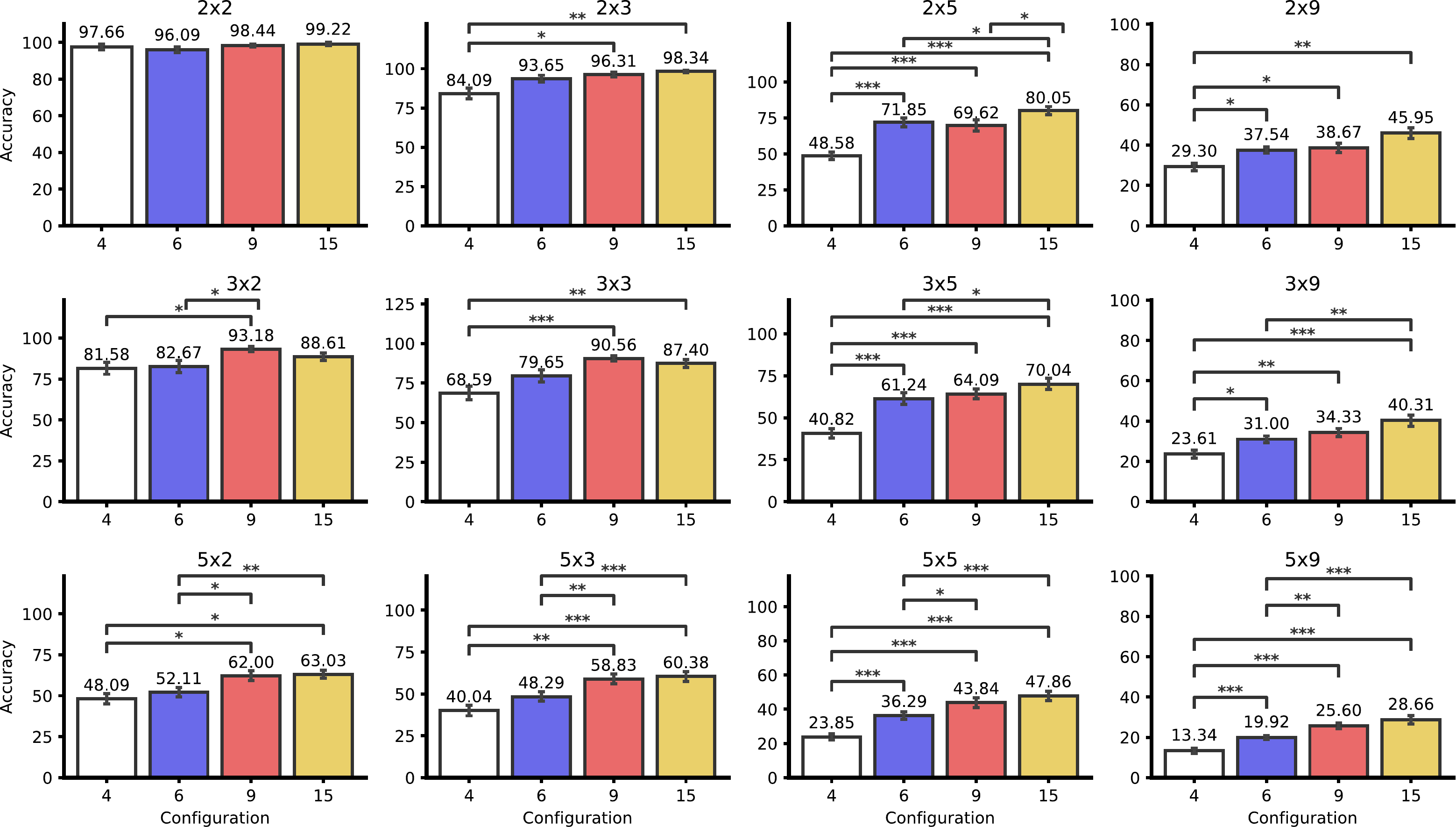}
  \caption{Localization accuracy of stationary sensations on the palm across different resolutions. Error bars are the standard errors. Results from post-hoc pairwise comparisons are shown (* $\leq$ 0.05, ** $\leq$ 0.01, *** $\leq$ 0.001)  }
  \label{fig:accuracy}
\end{figure*}

We analyzed the accuracy of localizing the target positions. To get a better overview, we further filtered the data to measure the accuracy of reducing from 5 columns to 3 and 2 equally spaced columns. Similarly, reducing from 9 rows to 5, 3, and 2 equally spaced rows. For each resolution, we conducted a RM-ANOVA with the factors \textsc{configuration} and \textsc{intensity}. The factor \textsc{intensity} did not have a significant effect on the accuracy of any of the resolutions. \autoref{tab:accuracy} summarizes the results of the RM-ANOVAs for the factor \textsc{configuration}. \autoref{fig:accuracy} visualizes the accuracy across varying resolution and the post-hoc tests.

\begin{table}
\centering
\resizebox{\linewidth}{!}{%
\begin{tabular}{cllccllll} 
\toprule
\multirow{2}{*}{\begin{tabular}[c]{@{}c@{}}Resolution \\(X*Y)\end{tabular}} & \multicolumn{1}{c}{\multirow{2}{*}{F}} & \multicolumn{1}{c}{\multirow{2}{*}{df}} & \multirow{2}{*}{p} & \multirow{2}{*}{$\eta_{g}^{2}$} & \multicolumn{4}{c}{\textsc{configuration} (EMM [lower CL, upper CL])}                                                              \\ 
\cline{6-9}
                                                                            & \multicolumn{1}{c}{}                   & \multicolumn{1}{c}{}                    &                    &                                         & \multicolumn{1}{c}{4}          & \multicolumn{1}{c}{6}          & \multicolumn{1}{c}{9}          & \multicolumn{1}{c}{15}          \\ 
\hline
2x2                                                                         & 1.00                                   & 2.54, 38.12                             & .393               & .041                                    & \emmcinb{97.7}{94.0}{100}{\%}  & \emmcinb{96.1}{92.1}{100}{\%}  & \emmcinb{98.4}{96.2}{100}{\%}  & \emmcinb{99.2}{97.6}{100}{\%}   \\
2x3                                                                         & 9.08                                   & 1.58, 23.70                             & .002               & .273                                    & \emmcinb{84.1}{76.6}{91.6}{\%} & \emmcinb{93.6}{88.6}{98.7}{\%} & \emmcinb{96.3}{93.1}{99.6}{\%} & \emmcinb{98.3}{96.4}{100}{\%}   \\
2x5                                                                         & 33.46                                  & 2.32, 34.80                             &  .001              & .455                                    & \emmcinb{48.6}{42.7}{54.5}{\%} & \emmcinb{71.8}{64.8}{78.9}{\%} & \emmcinb{69.6}{61.2}{78.0}{\%} & \emmcinb{80.0}{73.6}{86.5}{\%}  \\
2x9                                                                         & 11.09                                  & 2.64, 39.55                             &  .001              & .322                                    & \emmcinb{29.3}{25.3}{33.3}{\%} & \emmcinb{37.5}{34.1}{40.9}{\%} & \emmcinb{38.7}{33.9}{43.5}{\%} & \emmcinb{46.0}{39.8}{52.1}{\%}  \\
3x2                                                                         & 3.78                                   & 2.36, 35.41                             & .026               & .137                                    & \emmcinb{81.6}{73.5}{89.7}{\%} & \emmcinb{82.7}{74.7}{90.7}{\%} & \emmcinb{93.2}{89.5}{96.9}{\%} & \emmcinb{88.6}{83.7}{93.5}{\%}  \\
3x3                                                                         & 11.47                                  & 2.23, 33.43                             &  .001              & .296                                    & \emmcinb{68.6}{59.1}{78.1}{\%} & \emmcinb{79.6}{71.2}{88.1}{\%} & \emmcinb{90.6}{86.6}{94.6}{\%} & \emmcinb{87.4}{82.1}{92.7}{\%}  \\
3x5                                                                         & 31.50                                  & 2.68, 40.16                             &  .001              & .441                                    & \emmcinb{40.8}{35.1}{46.5}{\%} & \emmcinb{61.2}{53.7}{68.8}{\%} & \emmcinb{64.1}{57.4}{70.7}{\%} & \emmcinb{70.0}{62.8}{77.3}{\%}  \\
3x9                                                                         & 14.63                                  & 2.37, 35.61                             &  .001              & .332                                    & \emmcinb{23.6}{19.6}{27.6}{\%} & \emmcinb{31.0}{27.3}{34.7}{\%} & \emmcinb{34.3}{29.8}{38.8}{\%} & \emmcinb{40.3}{34.1}{46.5}{\%}  \\
5x2                                                                         & 9.12                                   & 2.39, 35.78                             &  .001              & .237                                    & \emmcinb{48.1}{41.5}{54.7}{\%} & \emmcinb{52.1}{45.7}{58.5}{\%} & \emmcinb{62.0}{55.5}{68.5}{\%} & \emmcinb{63.0}{57.5}{68.6}{\%}  \\
5x3                                                                         & 17.74                                  & 2.09, 31.35                             &  .001              & .335                                    & \emmcinb{40.0}{33.2}{46.9}{\%} & \emmcinb{48.3}{42.0}{54.6}{\%} & \emmcinb{58.8}{52.5}{65.2}{\%} & \emmcinb{60.4}{54.2}{66.5}{\%}  \\
5x5                                                                         & 31.79                                  & 2.40, 36.05                             &  .001              & .463                                    & \emmcinb{23.9}{19.8}{27.9}{\%} & \emmcinb{36.3}{31.7}{40.8}{\%} & \emmcinb{43.8}{37.2}{50.5}{\%} & \emmcinb{47.9}{41.9}{53.9}{\%}  \\
5x9                                                                         & 32.82                                  & 2.02, 30.35                             &  .001              & .481                                    & \emmcinb{13.3}{10.7}{16.0}{\%} & \emmcinb{19.9}{17.9}{22.0}{\%} & \emmcinb{25.6}{22.1}{29.1}{\%} & \emmcinb{28.7}{24.0}{33.3}{\%}  \\
\bottomrule
\end{tabular}
}
\caption{Results of RM-ANOVAs for the factor \textsc{configuration} on the dependent variable accuracy. }
\label{tab:accuracy}
\end{table}

\begin{figure*}
  \centering
  \includegraphics[width=\textwidth]{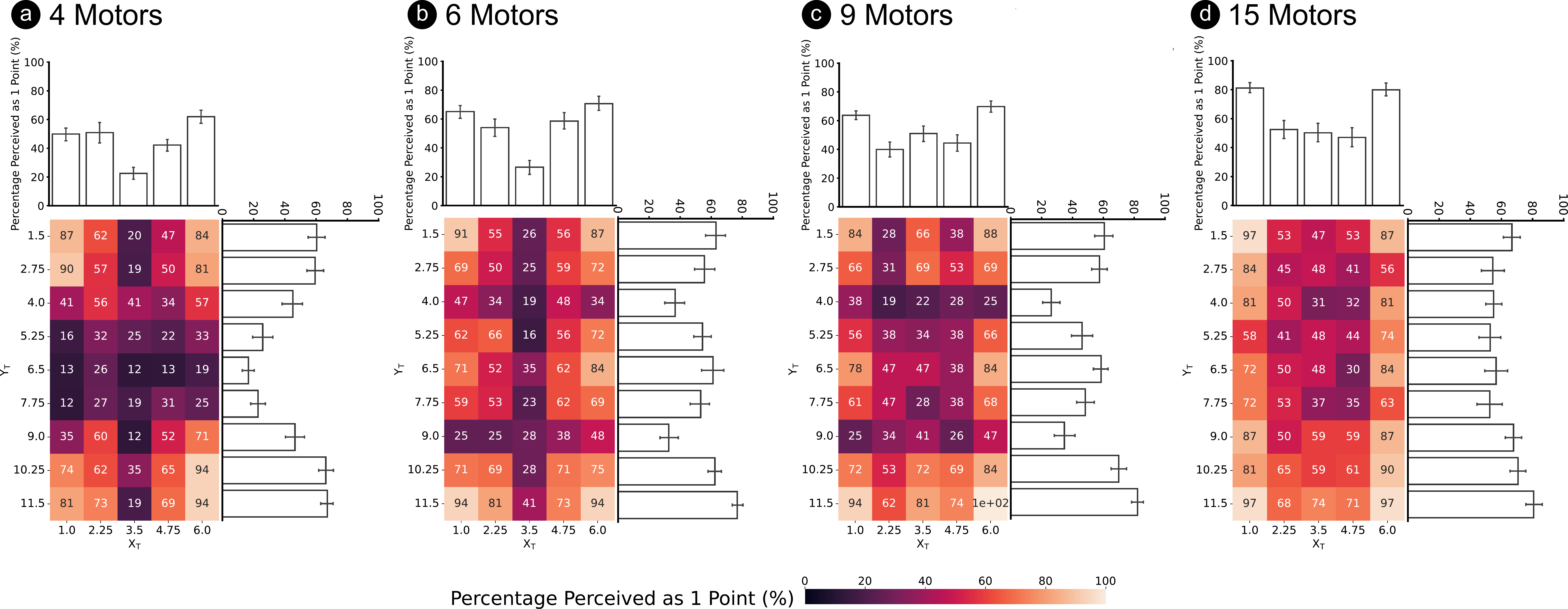}
  \caption{Probability of perceiving a single point for the factors \textsc{configuration} (a = \textsc{4 motor}, b = \textsc{6 motor}, c = \textsc{9 motor}, and d = \textsc{15 motor}), \Xt, and \Yt. Error bars are the standard errors.}
  \label{fig:pp}
\end{figure*}

\subsection{Number of Perceived Points}
We analyzed the binary dependent variable number of perceived points. Therefore, we used a logistic mixed model, estimated with ML and BOBYQA optimizer, with the fixed effects \textsc{configuration}, \textsc{intensity}, \Xt, and \Yt while including the participant as a random effect. The explanatory power~\cite{nakagawa} of the model was substantial $R^2 = 0.31$ and the part related to the fixed effects alone -marginal
$R^2$- was 0.18. We computed the 95\% confidence intervals and p-values using the Wald approximation. \autoref{fig:pp} visualizes the results.

\subsubsection{\textsc{configuration}}
\textbf{[MF-9]} The analysis revealed a significant \wald{3}{100.47}{< .001} main effect of the factor \textsc{configuration} on the number of perceived points. The \textsc{15 motor configuration} resulted in the highest probability \emmci{65.7}{56.3}{74.1}{\%} of perceiving vibration at a single location, followed by the \textsc{6 motor configuration} \emmci{56.8}{46.9}{66.2}{\%}, the \textsc{9 motor configuration} \emmci{55.2}{45.3}{64.7}{\%}, and finally the \textsc{4 motor configuration} \emmci{44.5}{35.1}{54.5}{\%}. Post-hoc pairwise contrasts confirmed all differences to be significant (p~<~.001) except between the \textsc{6} and \textsc{9 motor configurations}.

\subsubsection{\textsc{intensity}}
\textbf{[MF-10]} The analysis showed a significant \wald{1}{14.75}{< .001} main effect of the factor \textsc{intensity} on the number of perceived points between \textsc{0.5 intensity} \emmci{58.6}{48.9}{67.6}{\%} and \textsc{1.0 intensity} \emmci{52.8}{43.1}{62.3}{\%}.

\subsubsection{\Xt}
\textbf{[MF-11]} The analysis showed a significant \wald{4}{372.676}{< .001} main effect of the factor \Xt{} on the number of perceived points. The right \Xt = 6.0 \emmci{74.6}{66.2}{81.5}{\%} edge resulted in the highest probability of perceiving a single point, followed by the left \Xt = 1.0 \emmci{68.3}{59.0}{76.3}{\%} edge, and columns in the middle: \Xt = 2.25 \emmci{49.5}{39.7}{59.4}{\%}, \Xt = 4.75 \emmci{48.3}{38.5}{58.3}{\%}, and \Xt = 3.5 \emmci{35.2}{26.6}{44.8}{\%}. Post-hoc tests confirmed all pairwise contrasts as significant (p < .001) except the pair \Xt = 2.25 and \Xt = 4.75 (p~>~.05).

\subsubsection{\Yt}
\textbf{[MF-12]} The analysis showed a significant \wald{8}{359.388}{< .001} main effect of the factor \Yt{} on the number of perceived points. The bottom part of the grid resulted in the highest probability of perceiving a single point \Yt = 11.5 \emmci{81.2}{73.8}{86.9}{\%} and \Yt = 10.25 \emmci{71.2}{61.8}{79.0}{\%}. Followed by the rows at the top: \Yt = 1.5 \emmci{65.5}{55.6}{74.3}{\%} and \Yt = 2.75 \emmci{58.3}{47.9}{68.0}{\%}. Finally, we observed comparable performance for the rows in the middle: \Yt = 4.0 \emmci{38.9}{29.6}{49.2}{\%}, \Yt = 5.25 \emmci{43.9}{34.0}{54.3}{\%}, \Yt = 6.5 \emmci{48.0}{37.8}{58.4}{\%}, \Yt = 7.75 \emmci{43.0}{33.2}{53.4}{\%}, and \Yt = 9.0 \emmci{44.4}{34.4}{54.8}{\%}. Post-hoc tests confirmed significantly rising probability of one point perceived between the middle rows and the top rows (p < .001) and between the top rows and the bottom rows (p < .001).

\section{User Study 1: Discussion}
In the following, we discuss the findings from the first user study. 

\subsection{\textsc{configuration}}
\label{sec:config}
In general, localization accuracy increased as the number of actuators increased ([MF-1], [MF-4], [MF-7]). Observing spatial error metrics measured in our experiment, the usage of 9 actuators with a spacing of \SI{2.5}{\cm} along the width of the palm and \SI{5}{\cm} along the length is enough for accurate stationary sensations. Although euclidean distance was lowest using 15 actuators, the results of using 9 actuators were comparable, with no significant difference. Reducing to 6 actuators, however, results in a significant increase in euclidean errors (20\%) in comparison to 15 actuators ([MF-1]). 
Moreover, the additional column of actuators using 9 actuators decreases the X deviation significantly compared to configurations with a lower number of actuators, while being comparable to the configuration with 15 actuators ([MF-4]).
Although the Y deviation is significantly lower with 15 actuators compared to 9 actuators ([MF-7]), this only results in a significant decrease in localization accuracy for the $2\times5$ resolution (\autoref{fig:accuracy}). For all other resolutions, localization accuracy with 9 actuators was comparable to 15 actuators.

\begin{description}
    \item[INSIGHT-1] 9 actuators result in comparable localization performance of \textit{stationary} sensations as 15 actuators.
\end{description}

Regarding the perception of phantom sensations at a single location, results indicate that a \textsc{15 motor configuration} results in significantly higher probability of perceiving phantom sensations at a single location than all other \textsc{configurations} ([MF-9]). Although the probability of perceiving a single point is relatively low (65\%), this can be due to the fact that by asking the participants if they perceived a single point or multiple points, it is implied that multiple points can occur and participants are more inclined to say multiple points if they are in doubt. We calculated the probability of perceiving a single point with real sensations (always caused by a single actuator), and observed a probability of 75\% of users expressing that they felt a single point.

\begin{description}
    \item[INSIGHT-2] 15 actuators result in improved perception of phantom sensations at a single location. 
\end{description}

\subsection{\Xt{} \& \Yt{}}
We observed a systematic behaviour, where participants tend to localize vibrations closer to the edge of the device, leading to higher localization accuracies of locations at the edges and lower localization accuracies for locations in the middle ([MF-2], [MF-3], [MF-6], [MF-8]). A possible explanation for this behaviour is that with a lower number of actuators, phantom sensations are not perceived correctly at a single location and instead are perceived at the positions of the actuators generating them which are typically at the edges of the device (see \autoref{sec:config}, [\textbf{\textsc{insight-2}}]). We observed interaction effects supporting this ([IF-1], [IF-2]), where lower number of actuators show significant differences between locations at the edges and locations in the middle, and higher number of actuators show comparable performance across all locations -- indicating correct perception of phantom sensations at target location.

\begin{description}
    \item[INSIGHT-3] 9 and 15 actuators led to more accurate localization at target location. 4 and 6 actuators led to more frequent localization at the edges.  
\end{description}

Furthermore, results show that the X deviation values are considerably lower than the Y deviation values. This indicates that users could localize stationary sensations along the width of the palm more accurately than along the length. This is in line with prior work on localization accuracy of vibrotactile sensations on other body parts~\cite{10.1145/3432189}, where a transverse orientation resulted in better localization than a longitudinal orientation.

\begin{description}
    \item[INSIGHT-4] Localizing sensations along the width of the palm is more accurate than along the length.
\end{description}

Lastly, we analyzed users' ability to correctly localize stationary sensations with varying resolution. Observing the relationship between resolution and accuracy, it is clear that there is a trade-off between high accuracy and the number of distinct locations that can be localized as defined by the resolution. Typically, an application would require high accuracy (> 90\%) of localizing stationary sensations while still maintaining a reasonable resolution. A 3x3 resolution allows for identifying 9 different locations and maintains a high accuracy of localization (91\% for the \textsc{9 motor configuration}, see \autoref{fig:accuracy}).

\begin{description}
    \item[INSIGHT-5] A $3\times3$ resolution can be accurately recognised.
\end{description}

\subsection{\textsc{intensity}}
\textsc{Intensity} of the vibrations does not seem to have an influence on participants' ability to localize stationary tactile sensations. However, we observed a significant increase in the probability of perceiving a single point when using a lower \textsc{intensity} ([MF-10]). This is due to the fact that by increasing the intensity of the vibration, the positions of the actuators become more pronounced which affects the perception of phantom sensations.

\begin{description}
    \item[INSIGHT-6] Lower intensity vibrations result in improved perception of phantom sensations at a single location. 
\end{description}

\subsection{Comparison Between Touch and Vibrotactile Stimulation}
A two-point discrimination threshold of approximately \SI{0.8}{\cm} was observed for touch on the palm~\cite{SAPT}. Our findings show that for successful two-point discrimination of vibrotactile stimulation, a spacing of approximately \SI{2.8}{\cm} is required. This is calculated based on the upper CL of the euclidean distance with 15 actuators multiplied by two, to account for localization error of two locations. These values deviate considerably, highlighting the importance of basing design decisions for vibrotactile displays on vibrotactile perception.

\begin{description}
    \item[INSIGHT-7] Vibrotactile sensitivity on the palm deviates considerably from touch.
\end{description}

\section{User Study 2: Moving Tactile Sensations}
Based on the results of the first user study, we conducted a second user study to investigate moving tactile sensations. In particular, we aimed to answer the following research questions:

\begin{figure*}
  \centering
  \includegraphics[width=\textwidth]{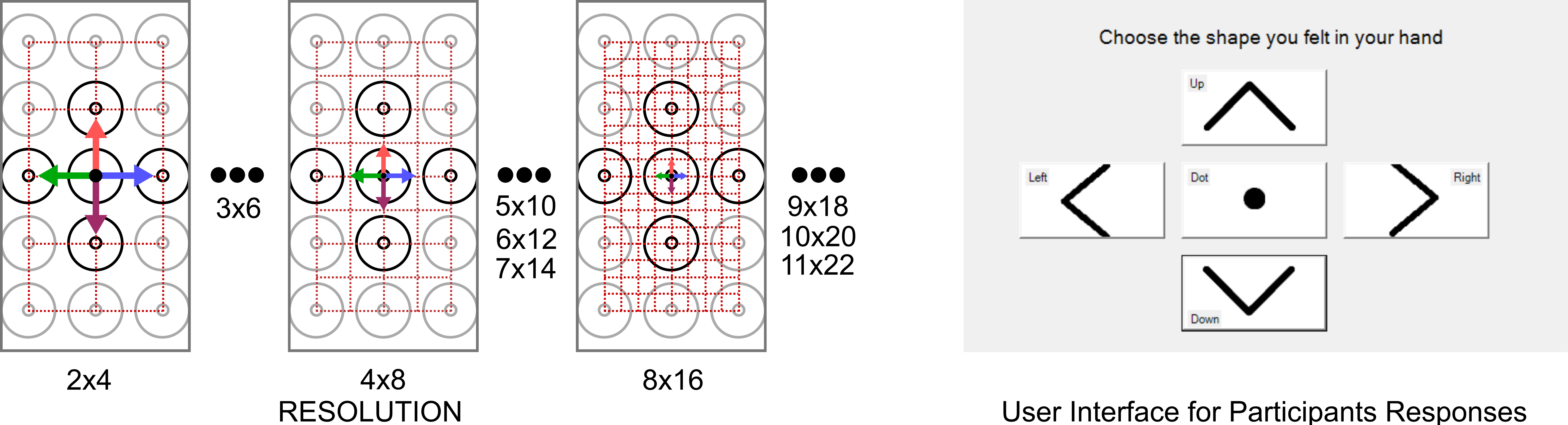}
  \caption{Study~2 evaluates 10~resolutions for \textit{moving} tactile sensations with a 15~motor configuration. Each \textit{moving} phantom sensation starts at the center motor and moves one step on the virtual resolution grid in one of four directions.}
  \label{fig:study2}  
  \label{fig:apparatus2}
\end{figure*}

\begin{description}
    \item[RQ1] How does the resolution effect recognition accuracy of users? Prior work explored a variety of resolutions for \textit{moving} sensations~\cite{7177686,5651759,6088602,6181408,6343763}. It is currently unclear, what the limits for accurate perception are.
    
    \item[RQ2] How does the resolution influence reaction time of users? We hypothesized that higher resolutions are more difficult to perceive and and hence lead to higher reaction times.
\end{description}

In the following, we detail on the study design, the procedure, apparatus, participants, dependent variables, and data analysis methods.

\subsection{User Study Design}
We varied the \textsc{resolution} of the vibrotactile grid and the \textsc{direction} of the sensation in a within-subject study design. \autoref{fig:study2} illustrates these independent variables. \textsc{Resolution} had 10 levels: \textsc{$2\times4$}, \textsc{$3\times6$}, \textsc{$4\times8$}, \textsc{$5\times10$}, \textsc{$6\times12$}, \textsc{$7\times14$}, \textsc{$8\times16$}, \textsc{$9\times18$}, \textsc{$10\times20$}, and \textsc{$11\times22$}. We aimed to cover a wide spectrum in \textsc{resolution} for a better understanding of users' performance.
\textsc{Direction} had 4~levels: \textsc{right}, \textsc{left}, \textsc{up}, and \textsc{down}. All sensations started at the middle actuator as shown in \autoref{fig:study2} and moved a distance equivalent to one cell in the respective \textsc{resolution}. The speed of the sensation was fixed at 0.5cm/s. We informally tested the influence of speed in pilot tests and found no difference in perception in the range 0.1-2cm/s. Inter-actuator spacing was fixed at \SI{2.5}{\cm}, which resulted in robust 1D phantom sensations from extensive pilot tests and based on prior work~\cite{10.1007/978-3-319-42321-0_5}. 

This resulted in a total of 40 ($10\times4$) conditions. We used a $10\times10$ balanced latin square to counterbalance the variable \textsc{resolution}. For each \textsc{resolution}, we randomized the order of appearance of the variable \textsc{direction}. Each condition was repeated three times, resulting in a total of 120~trials per participant

\subsection{Procedure}
Similar to the first user study, participants were welcomed to the lab, given a brief overview of the task, asked to sign an informed consent, and to fill out a short demographic questionnaire.

Participants wore noise cancelling headphones playing white noise throughout the experiment. The vibrotactile grid was attached to their dominant hand. At the beginning, a few (< 10) test trials were conducted to familiarize the participant with the task that were not recorded. Afterwards, participants began the experiment. The task was to feel a vibration and indicate if it was moving and in which direction. The participants experienced sensations that moved in the \textsc{directions up}, \textsc{down}, \textsc{right}, and \textsc{left}. However, an additional response was possible of feeling a point (\autoref{fig:apparatus2}) to indicate if they could not determine the direction.

Participants took a break (approximately five minutes long) after completing five different \textsc{resolutions}. In total, the experiment lasted about 20 minutes.

\subsection{Apparatus}
We used the \textsc{15 motor configuration} (for better perception of phantom sensations) throughout the experiment with the same hardware as the first user study. A pop-up window after each sensation was displayed to the users for inputting their response as shown in \autoref{fig:apparatus2}.

\subsection{Dependent Variables}
We recorded the following dependent variables:

\begin{description}
    \item[Accuracy:] whether the correct direction was identified. 
    \item[X Accuracy:] whether the correct direction was identified for left and right.
    
    \item[Y Accuracy:] whether the correct direction was identified for up and down.
    
    \item[Task Completion Time (TCT):] the time between the end of the vibration and the user's response.
    
\end{description}

\subsection{Participants}
We recruited 20 participants (15 male, 4 female, and 1 identified as gender variant), aged between 21 and 32 years old ($\mu = 23.75$, $\sigma = 2.69$). 19 of our participants were right-handed and one participant was left-handed. Participants reported no prior haptic feedback experience beyond the use of everyday smartphones and game controllers. All our participants reported no sensory processing disorders. 

\subsection{Data Analysis}
For the analysis of the binary dependent variable accuracy, we used a logistic mixed model, estimated with ML and BOBYQA optimizer, with the fixed effects \textsc{resolution} and \textsc{direction}, and including the participant as a random effect. For computing the confidence intervals and the p-values, we used the Wald approximation.

For the analysis of the continuous variable task completion time, we used a two-way RM-ANOVA with the factors \textsc{resolution} and \textsc{direction}, to compute the F-score and p-value of main and interaction effects. We used the same method for spherecity correction (Greenhouse–Geisser) and report the generalized eta-squared as in the first user study. For post-hoc tests, we used pairwise t-tests with Tukey adjustment. 

Similar to the first user study, we report the EMMs with the 95\% confidence intervals.
\section{User Study 2: Results}
This section reports the results of the second user study investigating \textit{moving} tactile sensations. Key observations are labelled with \textbf{[MF-\#]} for main effects and \textbf{[IF-\#]} for interaction effects.

\begin{figure*}
  \centering
  \includegraphics[width=\textwidth]{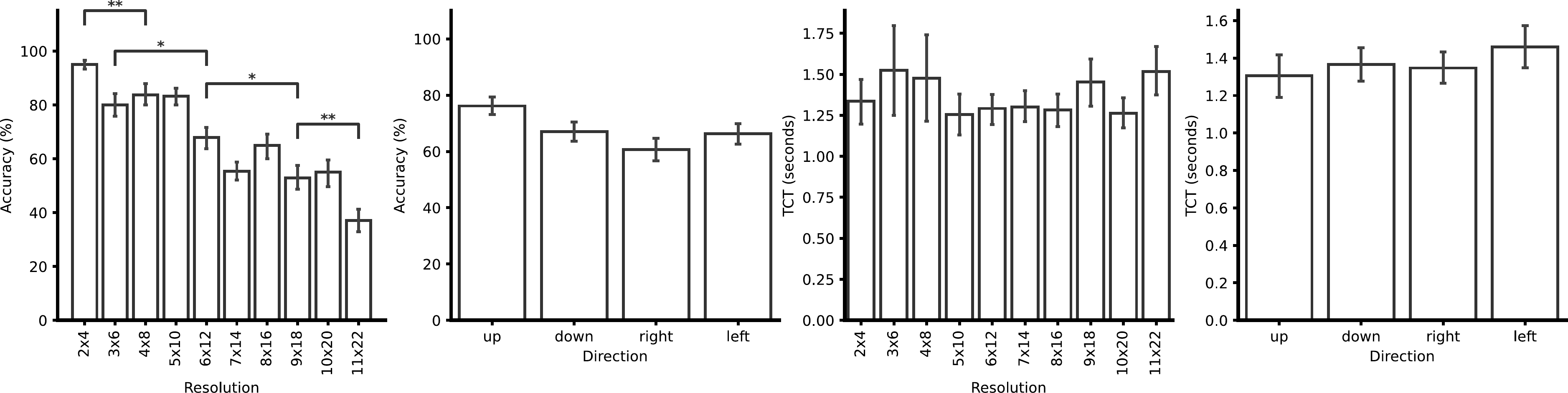}
  \caption{Accuracy and TCT for the factors \textsc{resolution} and \textsc{direction}. Error bars are the standard error. Results from post-hoc pairwise comparisons are shown (* $\leq$ 0.05, ** $\leq$ 0.01).}
  \label{fig:results2}
\end{figure*}

\subsection{Accuracy}
We analyzed the accuracy of users in identifying the direction of tactile sensation on the palm. \autoref{fig:results2} illustrates the results. The explanatory power of the model was substantial $R^2 = 0.34$ and the part related to the fixed effects alone (marginal $R^2$) was 0.27.

\subsubsection{\textsc{Resolution}}
\textbf{[MF-13]} Our analysis revealed a significant \wald{9}{86.96}{< .001} main effect of the factor \textsc{resolution} on the probability of correctly perceiving the moving sensation. The highest accuracy was achieved by the \textsc{$2\times4$ resolution} \emmci{96.2}{92.4}{98.1}{\%}, followed by \textsc{$5\times10$} \emmci{85.4}{79.0}{90.1}{\%}, \textsc{$4\times8$} \emmci{85.3}{79.0}{90.0}{\%}, \textsc{$3\times6$} \emmci{83.2}{76.1}{88.5}{\%}, \textsc{$6\times12$} \emmci{69.7}{61.1}{77.2}{\%}, \textsc{8x16} \emmci{66.4}{57.5}{74.2}{\%}, \textsc{$10\times20$} \emmci{56.4}{47.0}{65.3}{\%}, \textsc{$7\times14$} \emmci{56.0}{46.8}{64.8}{\%}, \textsc{$9\times18$} \emmci{53.3}{44.1}{62.2}{\%}, and \textsc{$11\times22$} \emmci{35.6}{27.4}{44.7}{\%}. Post-hoc tests are summarized in \autoref{fig:results2}.

\subsubsection{\textsc{Direction}}
We could not find a significant main effect of the factor \textsc{direction} on accuracy \wald{3}{1.97}{> .05}. The accuracy was comparable between the \textsc{directions up} \emmci{79.9}{73.9}{84.7}{\%}, \textsc{down} \emmci{74.2}{66.6}{80.7}{\%}, \textsc{left} \emmci{71.3}{64.0}{77.7}{\%}, and \textsc{right} \emmci{64.8}{57.0}{71.8}{\%}.

\subsection{X Accuracy}
We analyzed the accuracy of users for the \textsc{directions right} and \textsc{left}. The explanatory power of the model was $R^2 = 0.38$ and the marginal $R^2$ was 0.26.

\subsubsection{\textsc{Resolution}}
\begin{figure*}
  \centering
  \includegraphics[width=\textwidth]{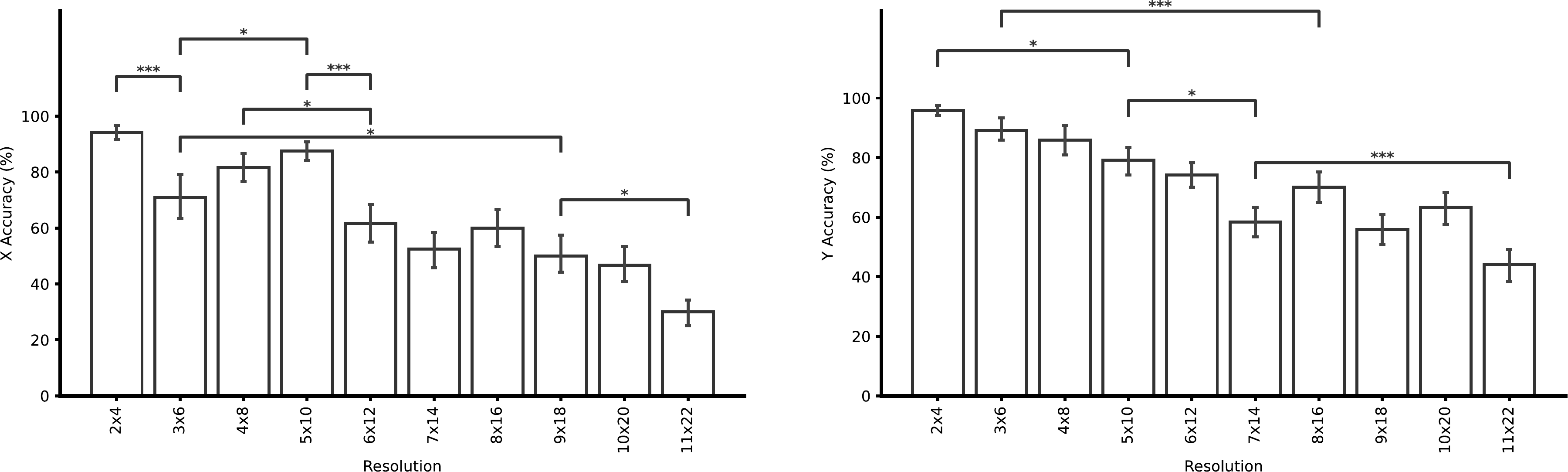}
  \caption{X accuracy and Y accuracy for the factor \textsc{resolution}. Error bars are the standard error. Results from post-hoc pairwise comparisons are shown (* $\leq$ 0.05, ** $\leq$ 0.01, *** $\leq$ 0.001).}
  \label{fig:results2xy}
\end{figure*}

\textbf{[MF-14]} Our analysis revealed a significant \wald{9}{79.67}{< .001} main effect of the factor \textsc{resolution} on the X accuracy. The \textsc{$2\times4$ resolution} resulted in the highest X accuracy \emmci{95.5}{90.0}{98.1}{\%}, followed by \textsc{$5\times10$} \emmci{89.9}{82.0}{94.6}{\%}, \textsc{$4\times8$} \emmci{84.3}{74.7}{90.8}{\%}, \textsc{$3\times6$} \emmci{73.4}{61.4}{82.6}{\%}, \textsc{$6\times12$} \emmci{63.5}{50.6}{74.7}{\%}, \textsc{$8\times16$} \emmci{61.6}{48.6}{73.1}{\%}, \textsc{$7\times14$} \emmci{53.1}{40.1}{65.7}{\%}, \textsc{$9\times18$} \emmci{50.1}{37.4}{62.9}{\%}, \textsc{$10\times20$} \emmci{46.3}{33.8}{59.3}{\%}, and \textsc{$11\times22$} \emmci{27.6}{18.1}{39.8}{\%}. \autoref{fig:results2xy} displays the result of post-hoc pairwise comparisons.

\subsubsection{\textsc{Direction}}
We could not find a significant effect of the factor \textsc{direction} on the X accuracy between \textsc{left} \emmci{72.3}{63.2}{79.8}{\%} and \textsc{right} \emmci{65.5}{55.7}{74.1}{\%} \wald{1}{0.16}{> .05}.

\subsection{Y Accuracy}
We analyzed the accuracy of users for the \textsc{directions up} and \textsc{down}. The explanatory power of the model was $R^2 = 0.33$ and the marginal $R^2$ was 0.26.

\subsubsection{\textsc{Resolution}}
\textbf{[MF-15]} Our analysis revealed a significant \wald{9}{85.25}{< .001} main effect of the factor \textsc{resolution} on the Y accuracy. The \textsc{$2\times4$ resolution} resulted in the highest Y accuracy \emmci{97.1}{91.4}{99.0}{\%}, followed by \textsc{$3\times6$} \emmci{90.3}{83.1}{94.6}{\%}, \textsc{$4\times8$} \emmci{87.2}{79.1}{92.4}{\%}, \textsc{$5\times10$} \emmci{80.9}{71.5}{87.7}{\%}, \textsc{$6\times12$} \emmci{75.8}{65.7}{83.7}{\%}, \textsc{$8\times16$} \emmci{71.3}{60.7}{79.9}{\%}, \textsc{$10\times20$} \emmci{65.7}{54.3}{75.6}{\%}, \textsc{$7\times14$} \emmci{59.0}{47.8}{69.3}{\%}, \textsc{$9\times18$} \emmci{56.4}{45.2}{67.0}{\%}, and \textsc{$11\times22$} \emmci{43.2}{32.4}{54.7}{\%}. \autoref{fig:results2xy} shows the result of post-hoc pairwise comparisons.

\subsubsection{\textsc{Direction}}
We could not find a significant effect of the factor \textsc{direction} on the X accuracy between \textsc{down} \emmci{74.2}{66.5}{80.6}{\%} and \textsc{up} \emmci{79.8}{73.9}{84.6}{\%} \wald{1}{1.66}{> .05}.

\subsection{Task Completion Time (TCT)}
We analyzed the time between the end of the stimulus and users' response.

\subsubsection{\textsc{Resolution}}
We could not find a significant main effect of the factor \textsc{resolution} on the TCT \statistics{3.25}{61.69}{0.52}{> .05}. TCT was comparable across \textsc{resolutions}. The highest TCT was observed for \textsc{$11\times22$} \emmci{1.52}{1.21}{1.83}{s} and \textsc{$3\times6$} \emmci{1.52}{0.92}{2.13}{s}, followed by \textsc{$4\times8$} \emmci{1.48}{0.91}{2.05}{s}, \textsc{$9\times18$} \emmci{1.45}{1.15}{1.76}{s}, \textsc{$2\times4$} \emmci{1.34}{1.05}{1.62}{s}, \textsc{$7\times14$} \emmci{1.30}{1.09}{1.49}{s}, \textsc{$6\times12$} \emmci{1.29}{1.10}{1.49}{s}, \textsc{$8\times16$} \emmci{1.28}{1.07}{1.50}{s}, \textsc{$10\times20$} \emmci{1.26}{1.06}{1.46}{s}, and \textsc{$5\times10$} \emmci{1.25}{0.99}{1.52}{s}.

\subsubsection{\textsc{Direction}}
We could not find a significant main effect of the factor \textsc{direction} on TCT \statistics{2.17}{41.30}{0.79}{> .05}. TCT was comparable for the \textsc{directions up} \emmci{1.31}{1.06}{1.55}{s}, \textsc{down} \emmci{1.37}{1.18}{1.56}{s}, \textsc{left} \emmci{1.46}{1.22}{1.70}{s}, and \textsc{right} \emmci{1.35}{1.16}{1.53}{s}.
\pagebreak
\section{User Study 2: Discussion}
In the following, we discuss the findings of our second user study.

\subsection{\textsc{resolution}}
In general, findings from our second user study show that more fine-grained resolutions are possible for \textit{moving} sensations in comparison to resolutions for vibrotactile point localization. This is consistent with related work on touch~\cite{SAPT}, where the localization of successive stimuli on the skin was better than simultaneous stimuli.

Our findings (MF-13) show that a $2\times4$ resolution results in 96.2\% correct recognition rate of the four directions \textsc{up}, \textsc{down}, \textsc{right}, and \textsc{left}. This is equivalent to a motion of \SI{2.5}{cm} on the palm. On the other hand, although a $5\times10$ resolution resulted in a significant decrease in recognition accuracy ([MF-13], [MF-15]) compared to a $2\times4$ resolution, we still observed a 85.4\% recognition rate of the four directions investigated in the user study. With this resolution, tactile motion equivalent to \SI{1.0}{\cm} can be accurately recognized. These results apply to tactile motion generated with 1D phantom sensations. While 1D phantom sensations are frequently used in the literature, further work is required for determining appropriate resolutions for 2D phantom sensations and other tactile illusions.

\begin{description}
    \item[INSIGHT-8] A $2\times4$ resolution results in high recognition accuracy (96.2\%).
    \item[INSIGHT-9] A $5\times10$ resolution enables higher resolution output while still maintaining high accuracy (85.4\%).
\end{description}

\subsection{\textsc{direction}}
Findings from the first user study unveiled a higher localization accuracy along the width of the palm in comparison to the length. This effect does not seem to apply to \textit{moving} sensations. We found comparable recognition rates for vertical and horizontal directions across \textsc{resolution}.

\subsection{Comparison Between Touch and Vibrotactile Stimulation}
Typically, spatial acuity is measured by two-point discrimination thresholds. However, experiments were also conducted that measured the perception of successive spatially distributed stimuli~\cite{SAPT} -- similar to motion on the skin. Results show that for touch on the palm, successive stimuli (\SI{0.5}{\cm} threshold) are discriminated more easily than simultaneous stimuli (\SI{0.8}{\cm} threshold). This trend is present in our results on vibrotactile stimulation, however, with considerably different absolute values (\SIrange{1}{2.5}{\cm} for \textit{moving} and \SI{2.8}{\cm} for \textit{stationary}). This is further confirmed by the high recognition accuracies we observed for \textit{moving} sensations with a length well below the localization accuracy of \textit{stationary} sensations.

\begin{description}
    \item[INSIGHT-10] Higher vibrotactile sensitivity was observed for \textit{moving} compared to \textit{stationary} sensations.
\end{description}

\section{Applications of Palm-based Tactile Displays}
\begin{figure*}
  \centering
  \includegraphics[width=\textwidth]{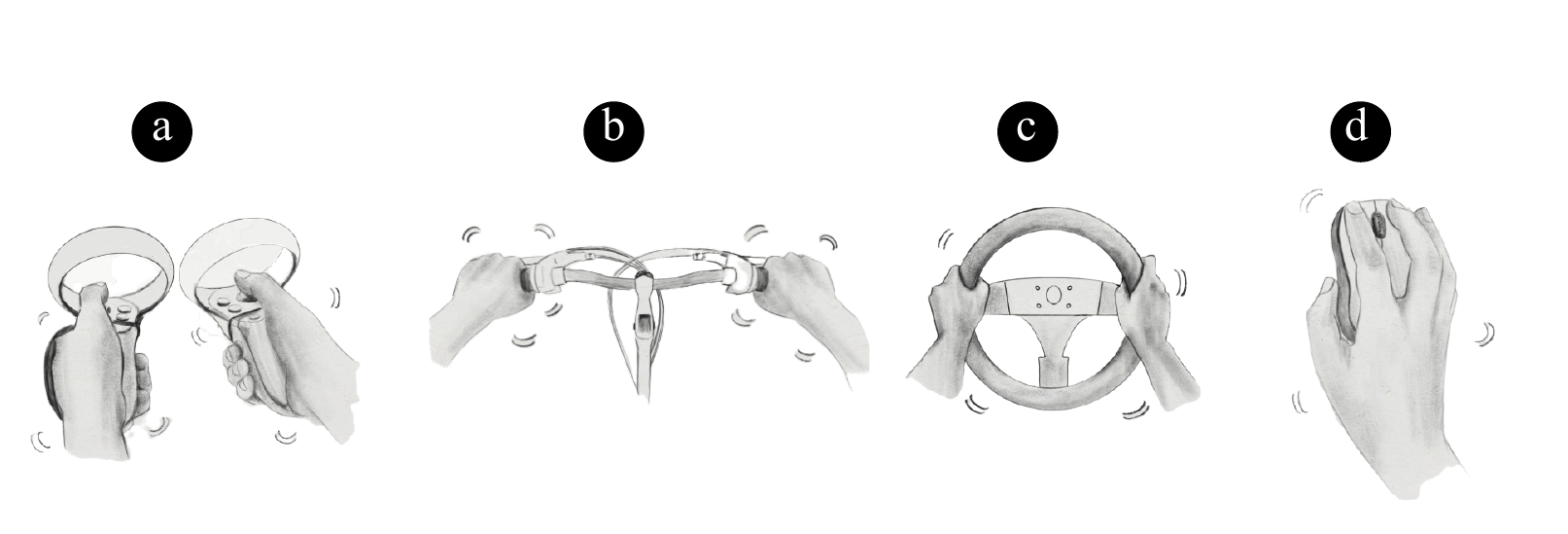}
  \caption{High-resolution palm-based tactile displays are applicable to many situations and can be embedded in a diverse set of devices, e.g., (a) VR controllers, (b) bicycle handlebar, (c) car steering wheel, and (d) computer mouse. }
  \label{fig:usecases}
\end{figure*}

\textit{Stationary} and \textit{moving} tactile sensations on the palm have many potential applications in the future (see \autoref{fig:usecases}).
These include high-resolution tactile feedback in video games and VR controllers to create more immersive experiences by simulating the exact contact points and shapes.
The handlebar of a bicycle and the steering wheel of a car could embed haptic actuators to give precise feedback of the surrounding (e.g., by communicating the exact location of other cars through vibrations) or navigation (e.g., by drawing the exact path the user needs to take on the user's palm through a \textit{moving} sensation).
Finally, traditional input devices such as mouse or pen input could be improved by haptic gestures that allow for a more \emph{expressive} and higher resolution tactile output.
These applications can benefit from the knowledge gained by our experiments by implementing the following design implications.

\section{Design Guidelines for Palm-based Tactile Displays}
In this section, we outline design guidelines derived based on the findings of our controlled user studies. These guidelines are aimed at two main interaction requirements: \emph{accurate} and \emph{expressive} sensations. An example application where \emph{accurate} interactions are required is pedestrian navigation~\cite{10.1145/2207676.2207734} or generally for \emph{accurate} communication of a \emph{discrete} set of instructions~\cite{Alvina:2015:OTC:2702123.2702341}. For these situations, \emph{accurate} sensations are required. 

On the other hand, more \emph{expressive} interaction is required for applications such as the communication of social touch~\cite{tasst} or generally for the communication of a more diverse set of \emph{continuous} sensations~\cite{Israr:2011:TBD:1978942.1979235} with more relaxed accuracy constraints. An example of this type of interaction in the context of \emph{stationary} sensations would be augmenting videos on handheld displays with vibrotactile feedback. An object appears on the scene and the user can feel it at the correct location on the hand. In this case, \emph{discrete} identification of a set of locations would not be the best approach. A tactile display with the ability to generate higher resolution sensations should be preferred. 

For \textit{stationary} vibrotactile sensations, we derive the following design guidlines (DG):

\begin{description}
    \item[DG-1] A smaller (approximately 2:1 ratio) inter-actuator spacing should be used along the width of the palm than the length. [\textbf{\textsc{insight-4}}]
    
    \item[DG-2] A $3\times3$ grid of points can be used for significantly \emph{accurate} interactions on the palm. [\textbf{\textsc{insight-5}}]
    
    \item[DG-3] Favor the use of a tactile display consisting of nine actuators with a spacing of \SI{2.5}{\cm} along the width and \SI{5}{\cm} along the length for \emph{accurate} interactions. [\textbf{\textsc{insight-1}}, \textbf{\textsc{insight-4}}]
    
    \item[DG-4] A high number (spacing $\leq \SI{2.5}{\cm}$) of actuators should be used for \emph{expressive} interactions. [\textbf{\textsc{insight-1}}, \textbf{\textsc{insight-2}}, \textbf{\textsc{insight-3}}]
    
    \item[DG-5] Lower intensity vibrations result in a higher probability of perceiving a phantom sensation at a single location. [\textbf{\textsc{insight-6}}]
\end{description}

For \textit{moving} tactile sensations we derive the following guidelines:

\begin{description}
    \item[DG-6] \emph{Accurate} interactions with \textit{moving} sensations can be achieved with a $2\times4$ resolution. [\textbf{\textsc{insight-8}}]
    
    \item[DG-7] A resolution of $5\times10$ maintains a reasonable recognition accuracy. This resolution can be used to generate more \emph{expressive} sensations. [\textbf{\textsc{insight-9}}]
\end{description}

\section{Limitations}
The design and results of our experiments impose some
limitations and directions for future work.

\subsection{Choice of Actuator}
Different vibrotactile actuators have been introduced and used by the literature, e.g. eccentric rotating mass (ERM)~\cite{10.1145/3432189} and linear resonant actuators (LRA)~\cite{eai}. These actuators differ in the mechanisms with which they produce vibrations on the skin. While LRAs vibrate perpendicular to the skin surface, ERMs rotate along the skin surface. We used an LRA actuator, known for its ability to produce localized sensations. Future work should investigate if and how much the choice of actuator affects perception.  

\subsection{Hand Pose}
The human hands are capable of a wide variety of poses and grips. We evaluated a flat hand pose in our user studies. Furthermore, our participants had to reduce the spacing between their fingers to make sure that actuators were in contact with the skin. While our results provide a valuable baseline for future research, further work is required to investigate how the hand pose affects perception.

\subsection{Interface Size}
We chose our prototype size based on prior work on back-of-device (smartphone) tactile interfaces and on the average hand size. While this allowed us to conduct our user studies with all our participants, our prototype covered only parts of the hand. The palm was always fully covered and depending on the hand size, the base of the fingers. Future devices should be tailored to users' hand sizes. 

\subsection{Real-World Applicability}
In our work, we investigated the perception of \textit{stationary} and \textit{moving} tactile sensations in a lab setting. We chose this approach to focus on the mere influence of the factors and to exclude external influences. While we are convinced that our results make a strong contribution to the future palm-based tactile displays, we also acknowledge that other settings might yield other results. Therefore, further work is necessary to understand how these results are transferable to in-the-wild settings.

\section{Conclusion}
This work explored the perception of vibrations on the palm. 
In a first user study, we evaluated the localization accuracy and perception of real and phantom \textit{stationary} sensations. Our findings show that a 9~actuator display with a $3\times3$ grid of points can be localized accurately. For better perception of phantom sensations, a display consisting of at least 15~actuators should be used combined with lower intensity vibrations. In a second controlled user study, we investigated the recognition accuracy of \textit{moving} tactile sensations. Findings show that a $2\times4$ resolution leads to accurately perceived \textit{moving} sensations, while a $5\times10$ resolution enables higher resolution output while maintaining high recognition rates. Based on the results, we derive and describe a set of design and usage guidelines for future palm-based tactile displays.

\bibliographystyle{ACM-Reference-Format}
\bibliography{palmvibe}

\end{document}